\documentclass[10pt,conference]{IEEEtran}
\usepackage{amsmath,amssymb}
\usepackage{amssymb}
\usepackage[pdftex]{graphicx,color}
\usepackage{psfig}
\usepackage{psfrag}
\usepackage{subfigure}
\usepackage{url}
\usepackage{array}
\usepackage{fontenc,times}
\usepackage{shadow,rotate}
\usepackage{cite}

\newtheorem{theorem}{\textbf{Theorem}}
\newtheorem{corollary}{\textbf{Corollary}}

\newtheorem{definition}{\textbf{Definition}}

\newtheorem{remark}{Remark}
\newcommand{\dv}{\mathbf} 
\newcommand{\mc}{\mathcal} 

\begin{document}
{\fontencoding{OT1}\fontsize{9.4}{11.25pt}\selectfont
\title{Multiaccess Channels with State Known to One Encoder: Another Case of Degraded Message Sets}
\author{Abdellatif Zaidi$\:^{\dagger}$ \qquad Shiva Prasad Kotagiri$\:^{\ddagger}$ \qquad J. Nicholas Laneman$\:^{\nmid}$ \qquad Luc Vandendorpe$\:^{\dagger}$\vspace{0.3cm}\\
$^{\dagger}$ \'{E}cole Polytechnique de Louvain, Universit\'{e} catholique de Louvain, LLN-1348, Belgium\\
$^{\ddagger}$ SERDES Technology Group, Xilinx Inc., San Jose, CA, USA-95124\\
$^{\nmid}$ Department of Electrical Engineering, University of Notre Dame, Notre Dame, IN-46556, USA\\
\{abdellatif.zaidi,luc.vandendorpe\}@uclouvain.be, skotagir@gmail.com, jnl@nd.edu\\
}

\maketitle

\begin{abstract}
We consider a two-user state-dependent multiaccess channel in which only one of the encoders is informed, non-causally, of the channel states. Two independent messages are transmitted: a common message transmitted by both the informed and uninformed encoders, and an individual message transmitted by only the uninformed encoder. We derive inner and outer bounds on the capacity region of this model in the discrete memoryless case as well as the Gaussian case. Further, we show that the bounds for the Gaussian case are tight in some special cases.
\end{abstract}

\section{Introduction}\label{secI}

We consider a memoryless multiaccess channel (MAC) controlled by random parameters, or channel states, as shown in Figure~\ref{ModelForMACwithAsymmetricCSI}. We assume that the channel state information (CSI) is known, non-causally, to only one of the two encoders, and the channel state is memoryless. Both encoders communicate with a decoder to which the channel states are not available. The informed encoder sends a common message and the uninformed encoder sends both the common message and an independent individual message. We refer to this situation as a case of degraded message sets \cite{KM77}. The decoder estimates the transmitted messages from the channel output.  In this paper, we study the capacity region of this communication model.

\begin{figure}[htpb]
\centering
\resizebox{0.9\linewidth}{!}{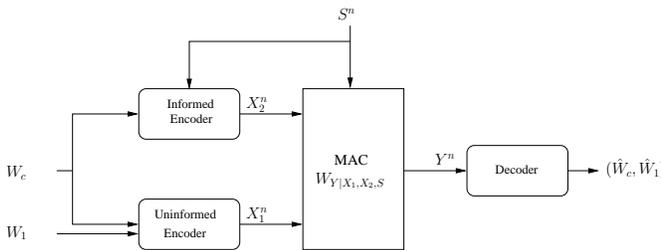}
\caption{Asymmetric state-dependent MAC with a degraded message set.}
 \label{ModelForMACwithAsymmetricCSI}
 \end{figure}

The study of channel models with random parameters initiated with Shannon \cite{Sh58} who derives the capacity of a single-user, state-dependent memoryless channel with causal transmitter state information. Gel'fand and Pinsker \cite{GP80} characterize the capacity of the same channel for the case in which the transmitter knows the channel states non-causally. Costa \cite{C83} extends Gel'fand and Pinsker's results to a Gaussian case and shows that dirty paper coding (DPC) completely mitigates the effect of the additive Gaussian channel state on the channel capacity. In part driven by a growing area of applications, the study of channels depending upon random parameters, especially multi-user models, has received considerable attention over the last decade. 
For a review on the subject of state-dependent channels and related work, the reader may refer to \cite{KSM08}.

The model shown in Figure~\ref{ModelForMACwithAsymmetricCSI} is an example of a state-dependent multiaccess model with {\it asymmetric} encoder CSI.  In \cite{KL04}, the authors introduce such asymmetry with state known to one, but not all, encoders. In \cite{PKEZ07}, the authors study asymmetry with different independent state components known to different encoders, focusing on lattice coding strategies in the Gaussian case. Incorporating degraded message sets, \cite{KL07,KL07a,SBSV07,SBSV07a} characterize the discrete memoryless capacity region for the case in which both encoders send a common message and the informed encoder sends, in addition, an independent individual message; \cite{SBSV07a} also provides an extensive treatment of the capacity region in the Gaussian case.

In this paper, we consider another case of degraded message sets for the MAC with asymmetric CSI, namely, the situation obtained by swapping the roles of the encoders in \cite{KL07,SBSV07a}. That is, the informed encoder sends a common message and the uninformed encoder sends the common message as well as an independent individual message. We derive inner and outer bounds on the capacity region of this model for both the discrete memoryless and Gaussian cases. For the Gaussian case, we also show that the bounds match in the case of strong Gaussian interference and high signal-to-noise ratio. 

Furthermore, specializing the results for the Gaussian case in this paper to the setup in which the uninformed encoder sends no individual message, we obtain the common message capacity established in \cite{SBSV07a}. Also, specializing the results to the case in which the informed encoder has no message to transmit, i.e., the \textit{helper problem} introduced in \cite{PKEZ07}, we obtain lower and upper bounds on the capacity of this model, for both discrete memoryless and Gaussian cases. For the Gaussian case, the bounds match in the limit of strong interference for some combinations of transmit powers and noise power. This same latter result is also established in \cite[Theorem 4]{PKEZ07} using lattice strategies. Finally, we mention that related scenarios of state-dependent relay channels with asymmetric CSI are studied in \cite{ZV07b,ZKLV08a,ZKLV08b,ZV09a,ZV09b,ZV09c}; and related scenarios of cognitive interference channels are studied in \cite{SBSV08,SVA-JS08,DMG08a,SE07a,MGKS08,LSBPSV07b}.

\section{Problem Setup}\label{secII}
We consider a stationary memoryless state-dependent MAC $W_{Y|X_1,X_2,S}$  whose output $Y \in \mc Y$ is controlled by the channel inputs $X_1 \in \mc X_1$ and $X_2 \in \mc X_2$ from the encoders and the channel state $S \in \mc S$ which is drawn according to a memoryless probability law $Q_S$. We assume that the channel state $S_i$ at time instant $i$ is non-causally known at one of the encoders, at the beginning of the transmission block. We refer to this encoder as "the informed encoder". The model shown in Figure ~\ref{ModelForMACwithAsymmetricCSI} is {\it asymmetric} in the sense that the channel state is known at only one of the two encoders. The informed encoder wants to send a common message $W_c$ and the uninformed encoder wants to send an independent individual message $W_1$ along with the common message $W_c$, in $n$ channel uses. Let $X_1^n=(X_{1,1},\cdots,X_{1,n})$ and $X_2^n=(X_{2,1},\cdots,X_{2,n})$ denote the inputs of uninformed encoder and the informed encoder, respectively, and $Y^n$ denotes the channel output. We note that the model considered in this paper has the uninformed encoder know the message of the informed encoder, and this is conceptually different from that considered in \cite{KL07,SBSV07a} in which it is the informed encoder who knows the message of the uninformed encoder.  

We assume that the common message $W_c$, with rate $R_c$, and the individual message $W_1$, with rate $R_1$, are independent random variables uniformly drawn from the sets $\mc W_c=\{1,\cdots,M_c\}$ and  $\mc W_1=\{1,\cdots,M_1\}$, respectively. For a positive-integer $n$ and a pair of non-negative real rate pair $(R_c,R_1)$, a $(2^{nR_c},2^{nR_1},n)-$code consists of two encoding functions  
$$\phi_1^n:\:{\mc W_c}{\times}{\mc W_1} \rightarrow \mc X_1^n \quad \text{and}\quad \phi_2^n:\:\:{\mc S^n}{\times}{\mc W_c} \rightarrow \mc X_2^n$$ at the uninformed encoder and the informed encoder, respectively, and a decoding function $\psi^n:\:\:\mc Y^n \rightarrow {\mc W_c}{\times}{\mc W_1}.$

From a $(2^{nR_c},2^{nR_1},n)-$code, the sequences $X_1^n$ and $X_2^n$ from the uninformed encoder and the informed encoder, respectively, are transmitted across a memoryless state-dependent MAC with conditional probability distribution
\begin{align}
&P_{Y^n|X_1^n,X_2^n,S^n}(y^n|x^n,\tilde{x}^n,s^n)=\nonumber\\
&\hspace{2cm}\prod_{i=1}^{n}W_{Y|X_1,X_2,S}(y_i|x_i,\tilde{x}_i,s_i).
\end{align}
The decoder estimates the messages from the channel output $Y^n$. The average probability of error is defined as $P_e^n:=\text{Pr}[\psi^n(Y^n) \neq (W_c,W_1)]$. 

A rate pair $(R_c,R_1)$ is said to be achievable if there exists a sequence of  $(2^{nR_c},2^{nR_1},n)-$codes $(\phi_1^n, \phi_2^n, \psi^n)$ with  $\lim_{n \rightarrow \infty} P_e^n=0$. The capacity region of the considered state-dependent MAC is defined as the closure of the set of achievable rate pairs.  

Due to space limitations, the results of this paper are either outlined only or mentioned without proofs. Detailed proofs can be found in \cite{ZKLV08d}.
\section{Discrete Memoryless Case}\label{secIII}
In this section, it is assumed that the alphabets $\mc S, \mc X_1, \mc X_2$ are finite.
\subsection{Inner Bound on the Capacity Region}\label{secIII_subsecA}

\begin{definition}
Let $\mc P$ denote the set of joint measures $P_{S,U_1,U_2,X_1,X_2,Y}$ of the form 
\begin{align}
&P_{S,U_1,U_2,X_1,X_2,Y}=\nonumber\\
&\hspace{1cm}Q_SP_{U_1}P_{X_1|U_1}P_{U_2|U_1,S}P_{X_2|U_1,U_2,S}W_{Y|X_1,X_2,S}
\label{MeasureForAchievableRateRegionDiscreteMemorylessChannel}
\end{align}
that satisfy
\begin{align}
I(U_2;Y|U_1,X_1)-I(U_2;S|U_1) > 0.
\label{Constraint__NonNegativeness__Probability__Measure}
\end{align}
\label{definition1}
\end{definition}
\vspace{-0.2cm}

\noindent The following theorem provides an inner bound on the capacity region of the state-dependent DM MAC shown in Figure~\ref{ModelForMACwithAsymmetricCSI}.

\begin{theorem}\label{TheoremAchievabeRateRegionDiscreteMemorylessChannel}The capacity region, $\mc C$, of the model shown in Figure \ref{ModelForMACwithAsymmetricCSI} contains the closure of the set of all rate-pairs $(R_c,R_1)$ satisfying 
\begin{align}
R_1 \: &< \: I(X_1;Y|U_1,U_2)\nonumber\\
R_1 \: &< \: I(X_1,U_2;Y|U_1)-I(U_2;S|U_1)\nonumber\\
R_c+ R_1 \: &< \: I(X_1,U_1,U_2;Y)-I(U_2;S|U_1),
\label{AchievableRateRegionDiscreteMemorylessChannel}
\end{align}
for some joint measure $P_{S,U_1,U_2,X_1,X_2,Y} \in \mc P$, where $U_1 \in \mc U_1$ and $U_2 \in \mc U_2$ are auxiliary random variables with
\begin{subequations}
\begin{align}
\label{BoundsOnCardinalityOfAuxiliaryRandonVariableU1ForAchievableRateRegionDiscreteMemorylessChannel}
&|\mc U_1| \leq |\mc S||\mc X_1||\mc X_2|+2\\
&|\mc U_2| \leq \Big(|\mc S||\mc X_1||\mc X_2|+2\Big)|\mc S||\mc X_1||\mc X_2|+2.
\label{BoundsOnCardinalityOfAuxiliaryRandonVariableU2ForAchievableRateRegionDiscreteMemorylessChannel}
\end{align}
\label{BoundsOnCardinalityOfAuxiliaryRandonVariablesForAchievableRateRegionDiscreteMemorylessChannel}
\end{subequations}
\end{theorem}
\vspace{-0.5cm}

First we generate a random codebook that we use to obtain the inner bound in Theorem \ref{TheoremAchievabeRateRegionDiscreteMemorylessChannel}. Next, we outline the encoding and decoding procedures. The corresponding error analysis can be found in \cite{ZKLV08d}. 

\noindent \textbf{Codebook Generation:} Fix a measure $P_{S,U_1,U_2,X_1,X_2,Y}$ satisfying \eqref{MeasureForAchievableRateRegionDiscreteMemorylessChannel}. Fix $\epsilon > 0$ and denote
\begin{align}
M_{c1}&= 2^{n[R_{c1}-2\epsilon]}\qquad & M_1&= 2^{n[R_1-4\epsilon]}\nonumber\\
M_{c2}&= 2^{n[R_{c2}-4\epsilon]}\qquad & J &= 2^{n[I(U_2;S|U_1)+2\epsilon]},
\label{ValuesForBinningVariablesInTheorem1}
\end{align}
for some $R_{c1} \geq 0$ and $R_{c2} \geq 0$.

\noindent The random encoders operate as follows. First, the uninformed encoder draws $M_{c1}$  i.i.d. vectors $\{\dv u_1(l_1)\}$, $l_1=1,\cdots,M_{c1}$, each with i.i.d. components drawn according to $P_{U_1}$. For each codeword $\dv u_1(l_1)$, the informed encoder draws a collection of $M_{c2}{\times}J$ auxiliary vectors $\{\dv u_2(l_1,l_2,j)\}$, $l_2=1,\cdots,M_{c2}$, $j=1,\cdots,J$, independently and each with i.i.d. components given $\dv u_1(l_1)$, according to the conditional measure $P_{U_2|U_1}$ induced by \eqref{MeasureForAchievableRateRegionDiscreteMemorylessChannel}. Also, for each codeword $\dv u_1(l_1)$, the uninformed encoder  draws $M_1$ i.i.d. vectors $\{\dv x_1(l_1,k)\}$, $k=1,\cdots,M_1$, each with i.i.d. components drawn according to the conditional measure $P_{X_1|U_1}$ induced by \eqref{MeasureForAchievableRateRegionDiscreteMemorylessChannel}.

\textbf{Encoding:} Suppose that a common message $W_c=l$ and an individual message $W_1=k$ are to be transmitted. We split the common message into two independent parts, $W_{c1}$ and $W_{c2}$. Let $(W_{c1},W_{c2})=(l_1,l_2)$. We denote by $R_{c1}$ and $R_{c2}$ the rates at which $W_{c1}$ and $W_{c2}$ are sent, respectively. The total rate for the common message is then $R_c=R_{c1}+R_{c2}$. 

\noindent The uninformed encoder transmits the vector $\dv x_1(l_1,k)$. Then, to transmit $l$, the informed encoder searches for the smallest $j$ such that $\dv u_2(l_1,l_2,j)$ is jointly typical with $(\dv u_1(l_1),\dv s)$. Denote this $j$ by $j^{\star}=j(\dv s,l_1,l_2)$. If such $j^{\star}$ is not found, or if the observed state is not typical, an error is declared and $j(\dv s,l_1,l_2)$ is set to $j=J$. Finally, the informed encoder transmits a vector $\dv x_2$ which is drawn i.i.d. conditionally given $\Big(\dv s, \dv u_1(l_1), \dv u_2(l_1,l_2,j^{\star})\Big)$ (using the conditional measure $P_{X_2|S,U_1,U_2}$ induced by  \eqref{MeasureForAchievableRateRegionDiscreteMemorylessChannel}).

\textbf{Decoding:} 
Upon observation of $\dv y$, the decoder at the receiver declares that $(\hat{l}_1,\hat{l}_2,\hat{k})$ is sent if there is a unique triple $(\hat{l}_1,\hat{l}_2,\hat{k})$ such that $\dv u_1(\hat{l}_1)$, $\dv u_2(\hat{l}_1,\hat{l}_2,j)$, $\dv x_1(\hat{l}_1,\hat{k})$ are jointly typical with $\dv y$, for some $j \in \{1,\hdots,J\}$. If there is no such triple or it is not unique, an error is declared. One can show that the receiver can decode reliably as long as $n$ is large and 
\begin{align}
R_1 \: &< \: I(X_1;Y|U_1,U_2) \nonumber\\
R_1 \: &< \: I(X_1,U_2;Y|U_1)-I(U_2;S|U_1)\nonumber\\
R_c+ R_1 \: &< \: I(X_1,U_1,U_2;Y)-I(U_2;S|U_1)\nonumber\\
R_{c2}+ R_1 \: &< \: I(X_1,U_2;Y|U_1)-I(U_2;S|U_1)\nonumber\\
R_{c2}\: &< \: I(U_2;Y|U_1,X_1)-I(U_2;S|U_1).
\label{BoundsForRateRegionsWithoutFourierMotzkinElimination}
\end{align}
Applying \textit{Fourier-Motzkin elimination} (see, e.g., \cite{L00}) to eliminate the variable $R_{c2}$ from \eqref{BoundsForRateRegionsWithoutFourierMotzkinElimination} yields the bounds \eqref{AchievableRateRegionDiscreteMemorylessChannel}, plus the additional constraint \eqref{Constraint__NonNegativeness__Probability__Measure}.


\subsection{Outer Bound on the Capacity Region}\label{secIII_subsecB}

The following theorem provides an outer bound on the capacity region of the state-dependent DM MAC shown in Figure~\ref{ModelForMACwithAsymmetricCSI}.

\begin{theorem}\label{TheoremOuterBoundDiscreteMemorylessChannel}
The capacity region, $\mc C$, of the model shown in Figure \ref{ModelForMACwithAsymmetricCSI} is contained in the closure of the set of all rate-pairs $(R_c,R_1)$ satisfying 
\begin{align}
R_1 \: &\leq \: I(X_1;Y|S,X_2), \nonumber\\
R_c+ R_1 \: &\leq \: I(X_1,X_2;Y|S)-I(X_1;S|Y),
\label{OuterBoundDiscreteMemorylessChannel}
\end{align}
for some probability distribution of the form
\begin{align}
&P_{S,X_1,X_2,Y}=Q_SP_{X_1}P_{X_2|X_1,S}W_{Y|X_1,X_2,S}.
\label{MeasureForOuterBoundDiscreteMemorylessChannel}
\end{align}\\
\end{theorem}

\textbf{Proof:} The proof of Theorem \ref{TheoremOuterBoundDiscreteMemorylessChannel} can be found in \cite{ZKLV08d}. 

We note that, through the subtracted term $I(X_1;S|Y)$, the outer bound \eqref{OuterBoundDiscreteMemorylessChannel} is tighter than that obtained by assuming that the channel state is available at both encoders and the decoder. Also, this outer bound is non-trivial and connects with a bounding technique that is used in \cite[Theorem 2]{SBSV07a}. However, in this work, this outer bound is proved using techniques that are different from those in \cite{SBSV07a}.
\section{The Gaussian MAC}\label{secV}
In this section, we consider a two-user state-dependent Gaussian MAC in which the channel states $S^n$ and the noise are additive and Gaussian. As in Section \ref{secIII}, we assume that only the second encoder knows the channel states, non-causally. The informed encoder sends only the common message, and the uninformed encoder sends both common and individual messages. 
\subsection{Channel Model}\label{secV_subsecA}
At time instant $i$, the channel output $Y_i$ is related to channel inputs $X_{1,i}$ and $X_{2,i}$ from the uninformed encoder and the informed encoder, respectively, and the channel state $S_i$ and the noise $Z_i$ by
\begin{align}
& Y_i=X_{1,i}+X_{2,i}+S_i+Z_i,
\label{ChannelModelForGaussianMACWithAsymmetricCSI}
\end{align}
where $S_i$ and $Z_i$ are zero-mean Gaussian random variables with variance $Q$ and $N$, respectively. The random variables $S_i$ and $Z_i$ at time instant $i \in \{1,\cdots,n\}$ are mutually independent, independent from the channel inputs $(X_1^n,X_2^n)$ and independent from $(S_j,Z_j)$ for $j \neq i$. 

We consider the individual power constraints on the transmitted power 
\begin{equation}
\sum_{i=1}^{n}X_{1,i}^2 \leq nP_1, \:\: \sum_{i=1}^{n}X_{2,i}^2 \leq nP_2.
\label{IndividualPowerConstraintsFullDuplexRegime}
\end{equation}
The definition of a code for this channel is the same as given in Section \ref{secII}, with the additional power constraint \eqref{IndividualPowerConstraintsFullDuplexRegime}.
\subsection{Lower and Upper Bounds on the Capacity Region}\label{secV_subsecB}

In this section, we establish inner and outer bounds on the capacity region of the state-dependent Gaussian MAC \eqref{ChannelModelForGaussianMACWithAsymmetricCSI}. The above results for the DM MAC can be readily extended to memoryless channels with discrete time and continuous alphabets using standard techniques \cite{G68}.  

The following theorem provides an achievable rate region.
\begin{theorem}\label{TheoremAchievableRateRegionGaussianChannel}
Let $\tilde{Q}(\rho) := (\sqrt{Q}+\rho\sqrt{P_2})^2$, for some $\rho \in [-1,0]$. The capacity region, $\mc C_{\text{G}}$, of the Gaussian model \eqref{ChannelModelForGaussianMACWithAsymmetricCSI} contains the union of the rate-pairs $(R_c,R_1)$ satisfying 
\begin{align}
R_1 \: &< \: \frac{1}{2}\log\Big(1+\frac{{\theta}P_1}{N+\frac{P_2\xi\tilde{Q}(\rho)(1-\alpha)^2}{P_2\xi+\alpha^2\tilde{Q}(\rho)}}\Big) \nonumber\\
R_1 \: &< \: \frac{1}{2}\log\Big(\frac{P_2\xi(P_2\xi+\tilde{Q}(\rho)+N+{\theta}P_1)}{P_2\xi\tilde{Q}(\rho)(1-\alpha)^2+N(P_2\xi+{\alpha^2}\tilde{Q}(\rho))}\Big)  \nonumber\\
R_c+ R_1 \: &< \frac{1}{2}\log\Big(1+\frac{(\sqrt{\bar{\theta}P_1}+\sqrt{1-\xi-\rho^2}\sqrt{P_2})^2+{\theta}P_1}{P_2\xi+\tilde{Q}(\rho)+N}\Big)\nonumber\\
&+ \frac{1}{2}\log\Big(\frac{P_2\xi(P_2\xi+\tilde{Q}(\rho)+N)}{P_2\xi\tilde{Q}(\rho)(1-\alpha)^2+N(P_2\xi+{\alpha^2}\tilde{Q}(\rho))}\Big),
\label{AchievableRateRegionGaussianChannel}
\end{align}
for some $\theta \in [0,1]$, $\bar{\theta}=1-\theta$, $\xi \in [0,1]$, $\rho \in [-\sqrt{1-\xi},0]$, and $\alpha \in \mathbb{R}$ is such that the logarithm terms on the RHS of \eqref{AchievableRateRegionGaussianChannel} are non-negative real.\\
\end{theorem}

\begin{remark}\label{remark1}
In Theorem~\ref{TheoremAchievableRateRegionGaussianChannel}, for the admissible values of $\alpha \in \mathbb{R}$ for fixed $\theta \in [0,1]$, $\xi \in [0,1]$ and  $\rho \in [-\sqrt{1-\xi},0]$, it is equivalent to consider those such that the logarithm term on the RHS of the second bound on $R_1$ and the second logarithm term on the RHS of the bound on $(R_c+R_1)$ are non-negative real. 
\end{remark}

\textbf{Proof:} A formal proof of Theorem \ref{TheoremAchievableRateRegionGaussianChannel} can be found in \cite{ZKLV08d}. The proof is based on the evaluation of the inner bound in Theorem \ref{TheoremAchievabeRateRegionDiscreteMemorylessChannel} using an appropriate jointly Gaussian distribution on $S, U_1, U_2, X_1, X_2$. Also, it is shown that the evaluation of the constraint \eqref{Constraint__NonNegativeness__Probability__Measure} gives the second logarithm term on the RHS of the bound on $(R_c+R_1)$ in Theorem~\ref{TheoremAchievableRateRegionGaussianChannel}; and this explains why this term should be non-negative real.

More informally, the uninformed encoder encodes the common message $W_c$ and the individual message $W_1$ in the same way as for a regular MAC. We decompose the input $X_1$ as
\begin{align}
X_1=U_1+\tilde{X}_1,
\label{InputOfUninformedEncoder}
\end{align}
where: $\tilde{X}_1$ is a zero-mean Gaussian with variance ${\theta}P_1$, is independent from $U_1$, $X_2$ and the state $S$; and $U_1$ is a zero-mean Gaussian with variance $\bar{\theta}P_1$, is correlated with $X_2$, with $\mathbb{E}[U_1X_2]=\rho'_{12}\sqrt{\bar{\theta}P_1P_2}$, and independent from $S$, for some $\theta \in [0,1]$, $\rho'_{12} \in [0,1]$. 

 \noindent The informed encoder encodes the common message by applying a DPC scheme as
\begin{align}
U_2=X_2-\frac{\sigma_{12}}{\bar{\theta}P_1}U_1+\Big(\alpha(1+\frac{\sigma_{2s}}{Q})-\frac{\sigma_{2s}}{Q}\Big)S,
\label{GDPCatInformedEncoder}
\end{align}
where $X_2$ is a zero-mean Gaussian with variance $P_2$ and is correlated with the channel state $S$ with $\mathbb{E}[X_2S]=\rho_{2s}\sqrt{P_2Q}$, for some $\rho_{2s} \in [-1,0]$; $\sigma_{12}=\mathbb{E}[X_1X_2]$ and $\sigma_{2s}=\mathbb{E}[X_2S]$ are the covariances, 
\begin{align}
\rho'_{12}=\frac{\sigma_{12}}{\sqrt{\bar{\theta}P_1P_2}},\quad \rho_{2s}=\frac{\sigma_{2s}}{\sqrt{P_2Q}},
\end{align}
and $\alpha$ is a scale parameter.

\noindent As it can be seen from the proof, different values of the $4$-tuple $(\theta,\rho'_{12},\rho_{2s},\alpha)$ satisfying 
\begin{equation}
{\theta}\bar{\theta}(1-\rho'^2_{12}-\rho^2_{2s}) \geq 0
\end{equation}
give different points that lie in the rate region defined by \eqref{AchievableRateRegionGaussianChannel}; and the union over all such $4$-tuples give the entire rate region in Theorem \ref{TheoremAchievableRateRegionGaussianChannel}. Also, the parameters $\rho$ and $\xi$ that appear in \eqref{AchievableRateRegionGaussianChannel} correspond to the substitution $\rho:=\rho_{2s}$ and $\xi:=1-\rho'^2_{12}-\rho^2_{2s}$.
 
The intuition for \eqref{GDPCatInformedEncoder} is as follows. The input of the informed encoder is composed of two parts. The first part, $\frac{\sigma_{12}}{\bar{\theta}P_1}U_1$, is correlated with the input of the uninformed encoder and, hence, it permits to obtain some {\it coherence} between the channel inputs for the transmission of the common message. The second part is generated using binning and is used to transmit additional common information and/or remove the effect of the interference on the communication, through a generalized DPC \cite{KL04,MS06}. 
\begin{remark}\label{remark2}
In the above coding scheme, we decomposed the input of the uninformed encoder into two independent parts, as shown by \eqref{InputOfUninformedEncoder}. Essentially, the part $U_1$ is used to obtain coherent transmission of the common message as we indicated previously, through correlation with $X_2$. As an alternative coding scheme, one can obtain this coherent transmission by splitting the input of the informed encoder, instead of that of the uninformed encoder. In this case, one part of the input of the informed encoder is correlated with the input of the uninformed encoder and is independent from the channel state (this part permits to transmit (part of) $W_c$ coherently), and the other is independent from the input of the uninformed encoder and is correlated with the channel state (this part permits to transmit additional common information and to combat the effect of the channel state). This idea is used for a related setup in \cite{ZKLV08a,ZKLV08b} in the context of relay channels with states at only the relay.
\end{remark}
We now derive an outer bound on the capacity of the state-dependent Gaussian MAC \eqref{ChannelModelForGaussianMACWithAsymmetricCSI}. 

\begin{theorem}\label{TheoremOuterBoundGaussianChannel}
 The capacity region, $\mc C_{\text{G}}$, of the Gaussian model \eqref{ChannelModelForGaussianMACWithAsymmetricCSI} is contained in the union of the rate-pairs $(R_c,R_1)$ satisfying
\begin{align}
R_1\: &\leq \:\frac{1}{2}\log\Big(1+\frac{P_1(1-\rho^2_{12}-\rho^2_{2s})}{N(1-\rho^2_{2s})}\Big)\nonumber\\
R_c+R_1\: & \leq \frac{1}{2}\log\Big(1+\frac{(\sqrt{P_1}+\rho_{12}\sqrt{P_2})^2}{P_2(1-\rho^2_{12}-\rho^2_{2s})+(\sqrt{Q}+\rho_{2s}\sqrt{P_2})^2+N}\Big)\nonumber\\
&+\frac{1}{2}\log\Big(1+\frac{P_2(1-\rho^2_{12}-\rho^2_{2s})}{N}\Big),
\label{OuterBoundGaussianChannel}
\end{align}
for some $\rho_{12} \in [0,1]$, $\rho_{2s} \in [-1,0]$ such that
\begin{equation}
\rho^2_{12}+\rho^2_{2s} \leq 1.
\label{AllowableCovarianceMatrixOuterBound}
\end{equation}\\
\end{theorem}
The proof of Theorem \ref{TheoremOuterBoundGaussianChannel} follows by evaluating the outer bound \eqref{OuterBoundDiscreteMemorylessChannel} using appropriate joint distribution of $S,X_1,X_2,S,Y$. Essentially, the proof is very similar to that of Theorem 6 in \cite{SBSV07a} and, for this reason, we omit it here. It is based on showing that for the Gaussian channel \eqref{ChannelModelForGaussianMACWithAsymmetricCSI}, one can restrict attention to jointly Gaussian $(S,X_1,X_2,Y)$ with $\mathbb{E}[X_1S]=0$, $\mathbb{E}[X_1X_2]=\sigma_{12}=\rho_{12}\sqrt{P_1P_2}$ and $\mathbb{E}[X_2S]=\sigma_{2s}=\rho_{2s}\sqrt{P_2Q}$. The values of the covariances $\sigma_{12}$ and $\sigma_{2s}$ are such that the covariance matrix $\Lambda_{S,X_1,X_2,Z}$ of $(S,X_1,X_2,Z)$ has a non-negative discriminant, i.e. $QP_1P_2N(1-\rho^2_{12}-\rho^2_{2s}) \geq 0$. For $Q > 0$, this implies that $\rho^2_{12}+\rho^2_{2s} \leq 1$. 

\subsection{Numerical Examples and Analysis of Some Special Cases}\label{secV_subsecC}
The inner bound in Theorem \ref{TheoremAchievableRateRegionGaussianChannel} and the outer bound in Theorem \ref{TheoremOuterBoundGaussianChannel} are plotted for two interesting cases ($P_1 > Q$ and $P_1 < Q$) in Figure \ref{Fig1IllustrativeExamples} and Figure ~\ref{Fig2IllustrativeExamples}. We observe that, for certain combinations of $(P_1,P_2,Q,N)$, the uninformed encoder fully benefits from the availability of the CSI at the informed encoder, and so transmits at its maximal rate, as if there were no interference.  Figure~\ref{Fig4IllustrativeExamples} shows the derived bounds for an example  combination of $(P_1,P_2,Q,N)$ for which the lack of knowledge of the interference at the first encoder causes an inevitable rate loss for this encoder (e.g., see the point $(0,R_1)$).

{\textbf{Special Cases:}} 
The bounds in Theorem \ref{TheoremAchievableRateRegionGaussianChannel} and Theorem \ref{TheoremOuterBoundGaussianChannel} match in some special cases.  

 1) Strong interference and high SNR. Investigating the bounds \eqref{AchievableRateRegionGaussianChannel} and \eqref{OuterBoundGaussianChannel} in the case of strong interference (i.e., $Q \rightarrow \infty$) and high $\text{SNR}_1=\frac{P_1}{N}$ and $\text{SNR}_2=\frac{P_2}{N}$, (i.e., $P_1,P_2 \gg N$), it can be easily shown that these bounds match in this case. This result is stated in the following corollary.
\begin{corollary}\label{Corollary1}
In the limit of strong interference, the capacity region of the state-dependent Gaussian MAC \eqref{ChannelModelForGaussianMACWithAsymmetricCSI} at high SNR is given by the union of the set of all rate pairs satisfying
\begin{align}
R_1\: &\leq \:\frac{1}{2}\log(1+\frac{P_1}{N})-o(1)\nonumber\\
R_c+R_1\: &\leq \:\frac{1}{2}\log(1+\frac{P_2}{N})-o(1),
\label{CapacityGaussianMACatStrongInterferenceAndHighSNR}
\end{align}
\end{corollary}
where $o(1) \rightarrow 0$ as $P_1,P_2 \rightarrow \infty$.

In \cite{PKEZ07}, the authors focus on lattice strategies to study a Gaussian MAC with independent additive Gaussian interferences known (non-causally) to different encoders. Among other results, the authors establish the capacity region of the Gaussian MAC with one Gaussian interference known to only one of the encoders and independent messages at the encoders, in the case of strong interference and  high SNR \cite[Theorem 5]{PKEZ07}. The above result in Corollary~\ref{Corollary1} can be viewed as a generalization of that in \cite[Theorem 5]{PKEZ07} to a case of degraded message sets at the encoders. 

 2) The uninformed encoder has no individual message to transmit. In this case, putting $\theta=0$ in \eqref{AchievableRateRegionGaussianChannel}, we easily see that the obtained bounds match, and so we get the expression of the {\it common message} capacity for the Gaussian case. The same common message capacity expression is derived in \cite{SBSV07}. We note that the common message capacity is studied, and derived, for the DM case as well in \cite{SBSV07}. 

 3) The informed encoder has no message to transmit. This model is introduced in \cite{PKEZ07} and is named the \textit{helper problem} therein. Based on lattice strategies, the authors derive the capacity region for the Gaussian model for the case of strong interference and $N \leq |P_1-P_2|$ \cite[Theorem 4]{PKEZ07}. By putting $R_c=0$ in the bounds established in this paper, we readily get lower and upper bounds on the capacity of the helper problem, for both DM and Gaussian cases. For the Gaussian case, the bounds are for general SNR and interference power. Furthermore, like in \cite{PKEZ07} the bounds for the Gaussian case match in the case of strong interference and  $N \leq |P_1-P_2|$. This same result, stated in the corollary below, is established in \cite[Theorem 5]{PKEZ07} using lattice strategies.
\begin{corollary}\label{Corollary2}
In \eqref{ChannelModelForGaussianMACWithAsymmetricCSI}, if $N \leq |P_1-P_2|$ and the informed encoder has no message to transmit, the capacity of the resulting helper problem in the limit of strong interference is given by 
\begin{align}
R_1=\frac{1}{2}\log(1+\frac{\min\{P_1,P_2\}}{N}).
\end{align}
\end{corollary}

\begin{figure}[!h]
        \begin{center}
        \includegraphics[width=\linewidth,height=0.6\linewidth]{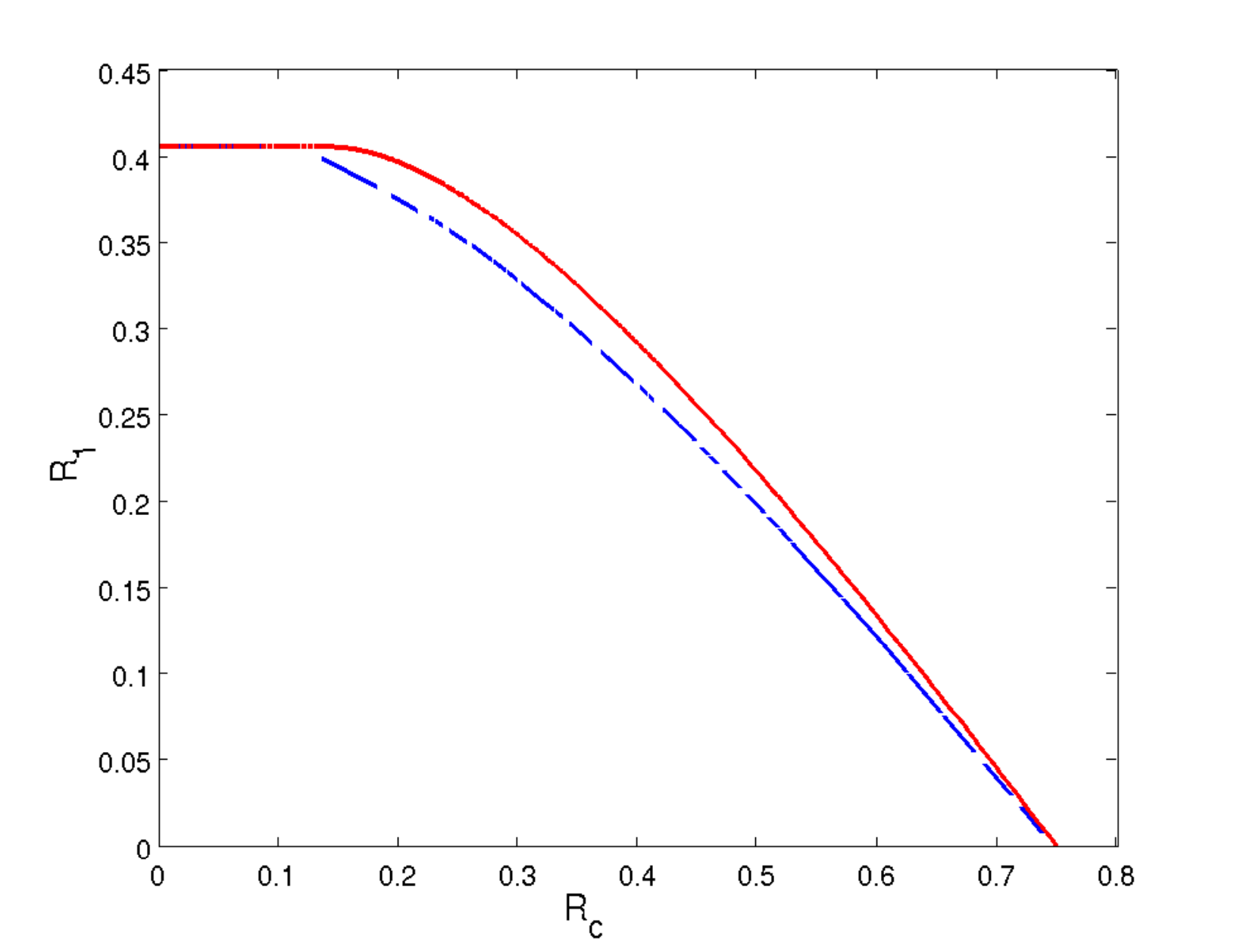}
        \caption{Inner (dashed) and outer (solid) bounds for $P_1=2.5$, $Q=1.5$, $P_2=N=2$.}
        \label{Fig1IllustrativeExamples}
        \end{center}
\end{figure}

\begin{figure}[!htpb]
\vspace{-0.5cm}
	\begin{center}
        \includegraphics[width=\linewidth,height=0.6\linewidth]{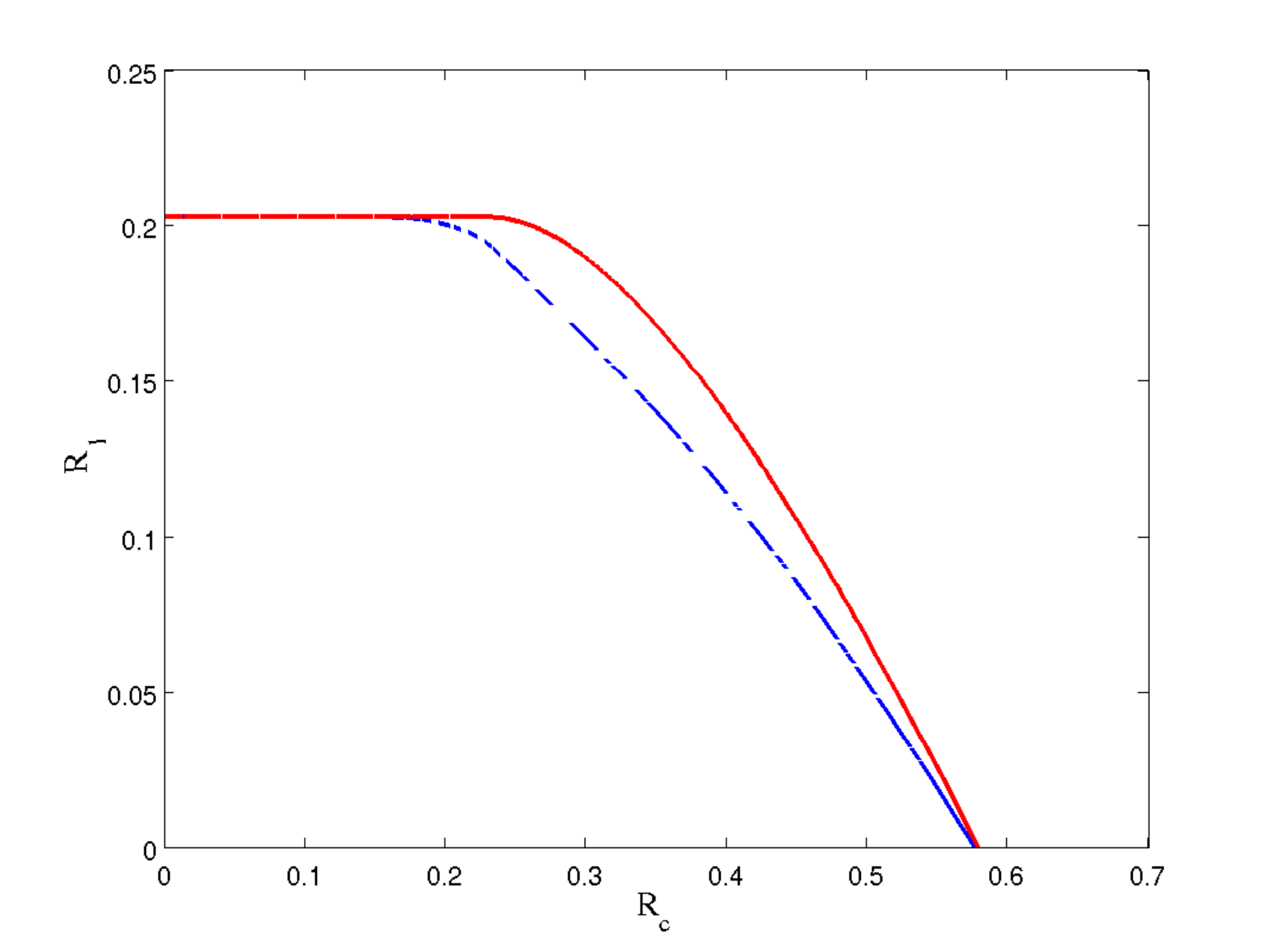}
 	\caption{Inner (dashed) and outer (solid) bounds for $P_1=1$, $Q=1.5$, $P_2=N=2$.}
        \label{Fig2IllustrativeExamples}
        \end{center}
\end{figure}
        
\begin{figure}[!htpb]
\vspace{-0.5cm}
	\begin{center}
        \includegraphics[width=\linewidth,height=0.6\linewidth]{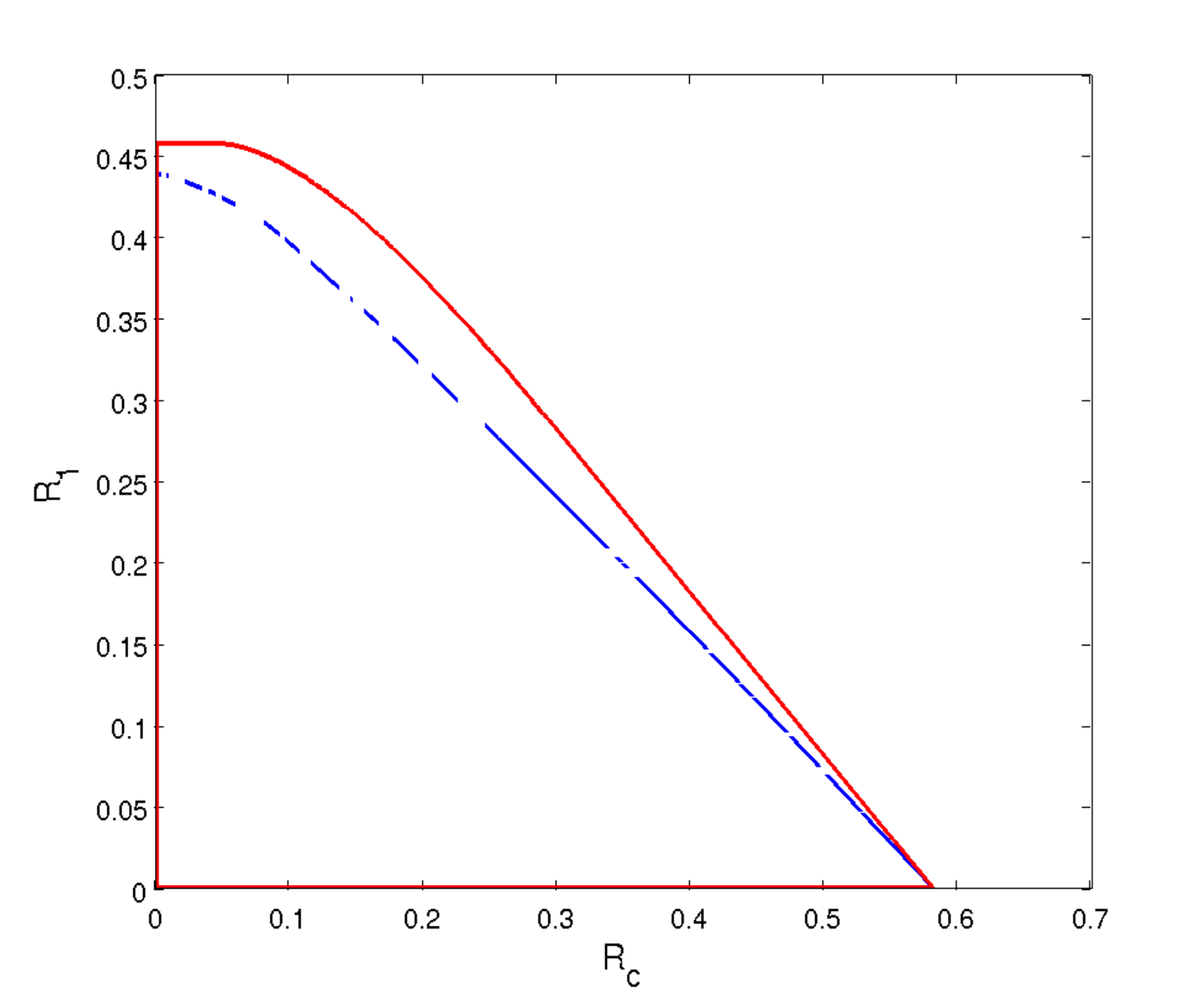}
        \caption{Inner (dashed) and outer (solid) bounds for $P_1=1.5$, $Q=2.5$, $P_2=N=1$.}
        \label{Fig4IllustrativeExamples}
	\end{center}
\vspace{-1cm}
\end{figure}

\bibliographystyle{IEEEtran}
\bibliography{paperISIT2009}
\end{document}